\documentclass{emulateapj}
\usepackage[]{natbib}
\usepackage{graphics}
     
\newcommand{\etal}{et al.}  
\newcommand{\per}{\ensuremath{^{-1}}}
\newcommand{\msun}{\ensuremath{{\rm M}_{\odot}}}
\newcommand{\kms}{km~s\ensuremath{^{-1}}} 
\newcommand{\mbh}{\ensuremath{M_\mathrm{BH}}}
\newcommand{\msigma}{\ensuremath{\mbh-\sigma_*}}
\newcommand{\sigmastar}{\ensuremath{\sigma_*}}

\shorttitle{LAMP: \msigma\ RELATION}
\shortauthors{WOO ET AL.}

\begin{document} 

\title{The Lick AGN Monitoring Project: the \msigma\ Relation for Reverberation-Mapped Active Galaxies}

\author{Jong-Hak Woo\altaffilmark{1,2,3}}

\author{Tommaso Treu\altaffilmark{4,5}}

\author{Aaron J. Barth\altaffilmark{6}}

\author{Shelley A. Wright\altaffilmark{2,7}}

\author{Jonelle L. Walsh\altaffilmark{6}}

\author{Misty C. Bentz\altaffilmark{2,6}}

\author{Paul Martini\altaffilmark{8}}

\author{Vardha N. Bennert\altaffilmark{4}}

\author{Gabriela Canalizo\altaffilmark{9}}

\author{Alexei V. Filippenko\altaffilmark{7}}

\author{Elinor Gates\altaffilmark{10}}

\author{Jenny E. Greene\altaffilmark{11,12}}

\author{Weidong Li\altaffilmark{7}}

\author{Matthew A. Malkan\altaffilmark{3}}

\author{Daniel Stern\altaffilmark{13}}

\author{Takeo Minezaki\altaffilmark{14}}

\altaffiltext{1}{Astronomy Program, Department of Physics and Astronomy, Seoul National University, Seoul, Korea, 151-742
({\tt woo@astro.snu.ac.kr}).}

\altaffiltext{2}{Hubble Fellow.}

\altaffiltext{3}{Department of Physics , University of California, Los Angeles, CA 90024.}

\altaffiltext{4}{Department of Physics, University of California,
Santa Barbara, CA 93106.}

\altaffiltext{5}{Sloan Fellow; Packard Fellow.}

\altaffiltext{6}{Department of Physics and Astronomy, 4129 Frederick
  Reines Hall, University of California, Irvine, CA 92697-4575.}

\altaffiltext{7}{Department of Astronomy, University of
  California, Berkeley, CA 94720-3411.}

\altaffiltext{8}{Department of Astronomy, and Center for Cosmology and
Astroparticle Physics, The Ohio State University, 140 West 18th Avenue, Columbus, OH 43210.}

\altaffiltext{9}{Department of Physics and Astronomy, University of California, Riverside, CA 92521, USA}

\altaffiltext{10}{Lick Observatory, P.O. Box 85, Mount Hamilton, CA 95140.}

\altaffiltext{11}{Department of Astrophysical Sciences, Princeton University, Princeton, NJ 08544, USA}

\altaffiltext{12}{Princeton-Carnegie Fellow.}

\altaffiltext{13}{Jet Propulsion Laboratory, California Institute of Technology, MS 169-527, 4800 Oak Grove Drive, Pasadena, CA 91109.}

\altaffiltext{14}{Institute of Astronomy, School of Science, University of Tokyo, 2-21-1 Osawa, Mitaka, Tokyo 181-0015, Japan.}

\begin{abstract}

To investigate the black hole mass vs. stellar velocity dispersion
(\msigma) relation of active galaxies, we measured the velocity
dispersions of a sample of local Seyfert 1 galaxies, for which we have
recently determined black hole masses using reverberation mapping. For
most objects, stellar velocity dispersions were measured from high 
signal-to-noise ratio optical spectra centered 
on the \ion{Ca}{2} triplet region ($\sim
8500$ \AA), obtained at the Keck, Palomar, and Lick Observatories.
For two objects, in which the \ion{Ca}{2} triplet region was
contaminated by nuclear emission, the measurement was based on high-quality 
$H$-band spectra obtained with the OH-Suppressing Infrared Imaging
Spectrograph at the Keck-II Telescope. Combining our new
measurements with data from the literature, we assemble a sample of 24
active galaxies with stellar velocity dispersions {\it and}
reverberation-based black hole mass measurements in the range of black
hole mass $10^{6}< \mbh/\msun < 10^{9}$.  We use this sample to
obtain reverberation mapping constraints on the slope and
intrinsic scatter of the \msigma\ relation of active
galaxies. Assuming a constant virial coefficient $f$ for the
reverberation mapping black hole masses, we find a slope
$\beta=3.55\pm0.60$ and the intrinsic scatter
$\sigma_{\rm int}=0.43\pm0.08$ dex in the relation $\log (M_{\rm BH} /
M_{\odot}) = \alpha + \beta \log (\sigma_{\ast} / {200}$ \kms), which
are consistent with those found for quiescent galaxies.  We derive an
updated value of the virial coefficient $f$ by finding the value which
places the reverberation masses in best agreement with the \msigma\
relation of quiescent galaxies; using the quiescent \msigma\ relation
determined by G\"ultekin et al.\ 
we find $\log f=0.72^{+0.09}_{-0.10}$ with an intrinsic scatter of
$0.44\pm0.07$ dex. 
No strong correlations between
$f$ and parameters connected to the physics of accretion (such as
the Eddington ratio or line-shape measurements) are found.  The
uncertainty of the virial coefficient remains one of the main sources
of the uncertainty in black hole mass determinations using
reverberation mapping, and therefore also in single-epoch spectroscopic
estimates of black hole masses in active galaxies.
\end{abstract}


\section{Introduction}

In the past decade, supermassive black holes have been found to be virtually
ubiquitous at the centers of local quiescent galaxies (for reviews, see
 Kormendy 2004; Ferrarese \& Ford 2005). Remarkably,
the mass of the central black hole (\mbh) correlates with the global
properties of their host spheroids, primarily the stellar velocity
dispersion ($\sigma_*$; Ferrarese et al. 2000; Gebhardt et al. 2000a). 
This \msigma\, relation is believed to be of fundamental
importance, providing one of the strongest empirical links between
galaxy formation/evolution and nuclear activity.  Measuring the
slope and scatter of the relation, and any possible dependence on
additional parameters, is essential to make progress in our
understanding of the co-evolution of galaxies and black holes.

One key open question is whether active galactic nuclei (AGNs) share
the same \msigma\ relation as quiescent galaxies, as would be expected
if the relation were universal and nuclear activity were just a random
transient phase.  Initial studies using several reverberation-mapped
Seyfert 1 galaxies show that the \msigma\ relation of active galaxies
is consistent with that of quiescent galaxies (Gebhardt et al. 2000b;
Ferrarese et al. 2001).  Other studies using larger samples with
single-epoch black hole masses suggest that the slope of the relation
may be different than that of quiescent galaxies, albeit at low
significance (e.g., Greene \& Ho 2006; Shen et al.\ 2008). However,
these later studies rely on an indirect estimator of black hole mass,
based on the kinematics and size of the broad-line region (BLR) as
inferred from single-epoch spectra. The method works as follows: the
broad-line profile gives a characteristic velocity scale $\Delta V$
(measured from either line dispersion, $\sigma_{\rm line}$, or the
full-width at half-maximum intensity, $V_{\rm FWHM}$); the average size 
of the BLR ($R_{\rm BLR}$) is determined from the empirical correlation with
optical luminosity (Wandel et al. 1999; Kaspi et al. 2005; Bentz et
al.\ 2009a); and the black hole mass is obtained as
\begin{equation}
    M_{\rm BH} = f \frac{(\Delta V)^2 R_\mathrm{BLR}}{G},
\label{eqn:mbh}
\end{equation}
where $G$ is the gravitational constant and $f$ is a virial
coefficient that depends on the kinematics and the geometry of the
BLR.  Although single-epoch mass estimates are believed
to be accurate to $\sim 0.5$ dex (e.g., Vestergaard \& Peterson 2006),
it is hard to quantify precisely their uncertainty and any possible
biases (Marconi et al. 2008; Onken et al. 2009). Thus, the slope and
intrinsic scatter of the \msigma\, relation of active galaxies remain
highly uncertain.

Multi-epoch data provide direct measurements of the BLR size via
reverberation mapping time lags ($R_{\rm BLR}= c\tau$, where $c$ is the
speed of light and $\tau$ is the measured reverberation time scale),
and more secure measurements of the broad-line kinematics as determined from the variable
component of the broad lines (Peterson et al.\ 2004).  Onken et al.\
(2004) combined reverberation black hole masses with measurements of
stellar velocity dispersion for 14 objects to measure the slope
of the \msigma\ relation, showing that AGNs and
quiescent galaxies lie on the same \msigma\, relation for an
appropriate choice of the virial coefficient ($ f
=5.5\pm1.8$).\footnote {Note that the virial coefficient, $\langle f
\rangle$, is $5.5$ when the line dispersion ($\sigma_{\rm line}$) is
used. If $V_{\rm FWHM}$ is used, $\langle f \rangle$ has to be
properly scaled depending on the $V_{\rm FWHM}/\sigma_{\rm line}$
ratio.  See Onken et al. (2004) and Collin et al. (2006) for details.}
Remarkably, the scatter in \mbh\ relative to the \msigma\ relation is
found to be less than a factor of 3. Unfortunately, their sample is very
small (14 objects) and covers a limited dynamic range ($6.2 < \rm log \mbh\ < 8.4$), 
making it difficult to simultaneously measure the intrinsic scatter
and the slope and investigate any trends with black hole mass, 
or with other properties 
of the nucleus or the host galaxy (cf. Collin et al.\ 2006).

Under the assumption that the relations should be the same for active
and quiescent galaxies, the Onken et al.\ (2004) study provides some
information on the geometry of the BLR, an absolute
normalization of reverberation black hole mass, and an upper limit to
the intrinsic scatter in the virial coefficient. Since the
reverberation-mapped AGN sample is the ``gold standard'' used to
calibrate all single-epoch mass estimates (e.g., Woo \& Urry 2002;
Vestergaard et al. 2002; McLure \& Jarvis 2002; Vestergaard \&
Peterson 2006; McGill et al.\ 2008; Shen et al. 2008), this comparison
is effectively a crucial link in establishing black hole masses for
all AGNs across the universe and for all evolutionary studies (e.g.,
McLure \& Dunlop 2004; Woo et al.\ 2006, 2008; Peng \etal\ 2006a,b;
Netzer \etal\ 2007; Bennert et al.\ 2010; Jahnke et al. 2009; Merloni
et al. 2010).

In this paper, we present new measurements of host-galaxy velocity
dispersion obtained from deep, high-resolution spectroscopy at the
Keck, Palomar, and Lick Observatories. The new measurements are
combined with existing data (e.g., Onken et al. 2004; Nelson et
al. 2004; Watson et al. 2008) and with recently determined black hole
masses from reverberation mapping (Bentz et al.\ 2009b) to construct
the \msigma\ relation of broad-lined AGNs for an enlarged sample
of 24 objects. For the first time, we are able to determine
{\it simultaneously} the intrinsic scatter and the slope of the \msigma\
relation of active galaxies. Also, by forcing the slope of the
\msigma\ relation to match that of quiescent galaxies, we determine
the average virial coefficient and its nonzero intrinsic
scatter. Finally, we study the residuals from the determined \msigma\
relation to investigate possible trends in virial coefficient with
properties of the active nucleus.

The paper is organized as follows. In \S 2, we describe the observations
and data reduction of the optical and the near-infrared (IR) data. New velocity
dispersion measurements are presented in \S 3, along with a summary of
previous measurements in the literature.  In \S 4, we investigate the
\msigma\ relation of the present-day AGNs, and determine the virial
coefficient in measuring \mbh.  Discussion and summary follow in \S 5
and \S 6, respectively.

\section{Observations}

The Lick AGN Monitoring Project (LAMP) was designed to determine the
reverberation time scales of a sample of 13 local Seyfert 1 galaxies,
particularly with low black hole masses ($<10^{7}$\msun).  NGC 5548,
the best-studied reverberation target, was included in our Lick
monitoring campaign to test the consistency of our results with
previous measurements.  The 64-night spectroscopic monitoring
campaign, along with nightly photometric monitoring, was carried out
using the 3-m Shane reflector at Lick Observatory and other, 
smaller telescopes.  The
detailed photometric and spectroscopic results are described by Walsh
et al.\ (2009) and Bentz et al.\ (2009b, 2010), respectively.  In
summary, nine objects including NGC 5548 showed enough variability in
the optical continuum and the H$\beta$ line to obtain the
reverberation time scales.  This significantly increases the size
of the reverberation sample, particularly for this low-mass range.

For the LAMP sample of 13 objects, we carried out 
spectroscopy using various telescopes, to measure stellar velocity
dispersions. Here, we describe each observation and the data-reduction
procedures.  The date, total exposure time, and other parameters for
each observation are listed in Table~\ref{observationstable}.
\subsection{Optical Data}

\begin{deluxetable*}{lcccccc}
\tablecaption{Observation Log} \tablehead{ \colhead{Galaxy} &
\colhead{Telescope/} & \colhead{UT Date} & \colhead{Exposure Time}
& \colhead{PA} & \colhead{Airmass} & \colhead{S/N} \\ \colhead{} &
\colhead{Instrument} & \colhead{} & \colhead{(s)} & \colhead{(deg)} &
\colhead{} & \colhead{} } 
\startdata 

Arp 151 & P200/DBSP & 2003-01-28 & 5400 & 210 & 1.10 & 79 \\

IC 1198 (Mrk 871) & Keck/ESI & 2008-03-02 & 900 & 285 & 1.25 & 87\\

IC 4218 & P200/DBSP & 2003-06-02 & 3600 & 5 & 1.23 & 61\\

MCG--06-30-15 & P200/DBSP & 2003-06-01 & 5400 & 10 & 2.70 & 112\\

Mrk 1310 & Keck/ESI & 2008-03-02 & 1800 & 323 & 1.19 & 84\\

Mrk 142 & Keck/OSIRIS & 2009-05-03 & 3600 & 0 & 1.19 & 17 \\

Mrk 202 & Keck/ESI & 2008-03-02 & 1200 & 204 & 1.32 & 73\\

Mrk 290 & Lick/Kast & 2008-04-13 & 3600 & 168 & 1.07 & 27 \\

NGC 4253 (Mrk 766) & P200/DBSP & 2001-06-26 & 1200 & 220 & 1.60 & 63 \\

                   & Keck/OSIRIS & 2009-05-03 & 3600 & 0 & 1.03 & 67 \\

NGC 4748 & Keck/ESI & 2004-02-17 & 1200 & 25 & 1.23 & 160 \\

         & Keck/OSIRIS & 2009-05-04 & 5400 & 0 & 1.37 & 61 \\ 

NGC 5548 & P200/DBSP & 2003-06-01 & 1800 & 59 & 1.06 & 100\\

NGC 6814 & P200/DBSP & 2003-06-01 & 3600 & 0 & 1.38 & 165 \\

SBS 1116+583A & Keck/ESI & 2008-03-02 & 1800 & 216 & 1.40 & 53 \\

\tablecomments{For OSIRIS observations, the position angle (PA) 
refers to the direction of the long axis of the IFU; for 
all other observations it refers to the slit PA.  
The S/N refers to the signal-to-noise ratio per pixel 
in the extracted spectra, at $\sim$8400-8700 \AA\ 
for the \ion{Ca}{2} triplet spectral region, or at 
$\sim$1.47--1.61 $\mu$m for the $H$-band spectra.}
\label{observationstable}
\end{deluxetable*}

\subsubsection{Palomar Observations}

Observations of six galaxies were obtained with the Double
Spectrograph \citep[DBSP;][]{og82} at the Palomar Hale 5-m telescope
(P200).
A 2\arcsec-wide slit was used, along with the D68 dichroic.  On the red side
of the spectrograph, we used a 1200 lines mm\per\ grating blazed at
9400 \AA, covering the wavelength range 8330--8960 \AA\ at an
instrumental dispersion of $\sigma_i \approx 30.4$ \kms.  

Each galaxy was observed with the slit oriented approximately at the
parallactic angle (Filippenko 1982) 
for the midpoint of the observation.  Typically two
or three exposures were taken for each galaxy to aid in cosmic-ray
removal.  Flux standards and a range of velocity template stars
(primarily K-type giants) were observed during each night. 

\subsubsection{Lick Observations}

Mrk 290 was observed with the Kast Double Spectrograph (Miller \& Stone 
1993) at the Shane 3-m telescope at Lick Observatory on 2008 April 13
(UT dates are used throughout this paper), 
during our AGN monitoring campaign.  For these observations, we
used the D55 dichroic and the 830/8460 grating on the red side of the
spectrograph, covering the wavelength range 7570--9620 \AA\ at a scale
of 1.7 \AA\ pixel$^{-1}$.  A 2\arcsec-wide slit was used and oriented at
the parallactic angle for the midpoint of the exposure sequence, and
the instrumental dispersion was $\sigma_i \approx 59$ \kms\ at 8600
\AA.  We obtained four 900 s exposures of Mrk 290 with this setup,
along with short exposures of three velocity template stars and the
flux standard star BD+28$^\circ$4211.

\subsubsection{Keck Observations}

Five galaxies were observed with the Echellete Spectrograph and Imager
\citep[ESI;][]{sheinis02} at the Keck-II 10-m telescope.  In ESI
echellette mode, the observations cover 3900--11000 \AA\ in 10
spectral orders at a scale of $\sim$11.4 km s\per\ pixel\per.  We used
a 0\farcs75-wide slit, giving an instrumental dispersion of $\sigma_i
\approx 22$ \kms.

The slit was oriented at the parallactic angle in all observations.
During twilight of each night, we observed flux standards and several
velocity template stars with spectral types ranging from G8III to K5III.

\subsubsection{Reductions}

We reduced the DBSP and Kast data using a series of IRAF\footnote{IRAF
is distributed by the National Optical Astronomy Observatories, which
are operated by the Association of Universities for Research in
Astronomy, Inc., under cooperative agreement with the National Science
Foundation.} scripts.  The standard spectroscopic data-reduction
procedure, including bias subtraction, flat fielding, wavelength
calibration, spectral extraction, and flux calibration was performed
for each data set.  Optimal extraction \citep{H86} was used for obtaining
one-dimensional spectra to achieve maximal signal-to-noise ratio (S/N) 
on the stellar features. The typical extraction radius was $\sim2-3''$, 
which corresponds to a physical radius of $\sim1$ kpc.

The ESI observations require some special preparation and calibration
before they can be combined and used for kinematic analysis. The
calibration process undertaken includes the following steps: bias
subtraction, flat-fielding, cosmic-ray rejection, wavelength
calibration, rectification, and sky subtraction. These steps were
performed by the IRAF package {\tt EASI2D}, which was developed by
David J.~Sand and Tommaso Treu \citep{sand04} for easy extraction of
echelle orders. Approximate flux calibration was performed using ESI
response curves measured during photometric nights. During wavelength
calibration the spectra were rebinned to uniform steps in
$\log(\lambda)$ corresponding to 11.4 km s$^{-1}$, which is close to
the native pixel scale of ESI, thus minimizing covariance between
pixels and loss of resolution due to rebinning. The root-mean square 
(rms) residuals in
the wavelength solution are much smaller than a pixel (typically 5\%)
and therefore negligible.  Spectral resolution was measured from night-sky 
lines and from wavelength-calibration lines taken in the same
configuration and reduced through the same pipeline to include all
instrumental effects. 
As for the DBSP and Kast data, optimal
extraction was used for obtaining one-dimensional spectra.  The
typical extraction radius of the ESI spectra was $\sim1-1.5''$, which
corresponds to a physical radius of $\sim0.5$ kpc at the mean redshift
of the observed AGNs.

\subsection{Near-Infrared Data}

\subsubsection{Keck Observations}

For three galaxies, we obtained $H$-band spectra using the
integral-field unit (IFU) OH-Suppressing Infrared Imaging Spectrograph
\citep[OSIRIS;][]{larkin06} at the Keck-II telescope, operated with the
laser guide-star adaptive optics (LGS-AO) system \citep{wiz06}.  Using
the high angular resolution in the AO-corrected data, our goal was to
extract the light from an annulus around the nucleus if necessary, by
excluding light from the central, AGN-dominated region, similar to what
was done by \citet{watson08}.

The OSIRIS observations were obtained on 2009 May 3 and  
4, during the first half of each night.  We used the $H$ broadband  
(Hbb) filter and the 0\farcs1 lenslet scale, giving a field of view of  
$1\farcs6\times6\farcs4$, with spectra covering the range 1473--1803  
nm at a spectral resolving power of $R\approx3800$.  The long spatial  
direction of the IFU was oriented north-south for all observations.

The observing sequence for each galaxy consisted of sets of four  
exposures. In three exposures the nucleus of the galaxy was  
placed on the IFU. Then one sky exposure with an offset of 21$\arcsec$  
to the southeast was taken. The three on-source exposures in each set were  
dithered northward by 0\farcs6 between exposures.  Each individual  
exposure had a duration of 600 s for NGC 4748, or 300 s for Mrk 142 and Mrk  
766, and total on-source exposure times for each source are listed in  
Table~\ref{observationstable}.  
Immediately preceding and following  
the observation sequence for each galaxy, we observed an A0V star for  
telluric correction and flux calibration. The A0V stars were observed  
using a sequence of two on-source and one off-source exposure, with  
typical exposure times of 5--10 s.  In order to obtain unsaturated  
exposures of these bright stars, the AO loops were not closed during  
the star observations. On each night, velocity template stars were  
also observed, following the same observing sequence as used for the  
telluric correction stars. We observed several K and M-type giant  
stars as velocity templates, and A0V telluric correction stars  
were also observed close in time and airmass to the velocity template  
stars.

The nucleus of each galaxy was used as a tip-tilt reference source
($V\approx15-16$ mag) for the LGS-AO system, and the laser was
propagated on-axis for each dithered exposure. It is difficult to
quantify the image quality or determine the Strehl ratio in the galaxy
exposures, since the galaxy nuclei are not point sources. However, the
typical LGS-AO performance for tip-tilt point sources for these
similar magnitudes yields $K$-band Strehl ratios of 40\% to 50\%
\citep{dam06,dam07}. Therefore, given that the quality of both nights
was excellent, we conservatively estimate the $H$-band Strehl ratio to
be $\sim$15\%--20\% for our galaxy observations.

Prior to this OSIRIS run, the OSIRIS detector temperature had been
rising from its previous operating temperature of 68 K.  During these
two nights, the detector temperature was steady at 74 K, which
resulted in a higher level ($\sim$40\%) of dark current than normal
for OSIRIS. The higher dark current did not significantly impact our
observations, since our exposures were not dark or background limited.
In the afternoon, multiple dark frames were taken with exposure times
matching those of our observations, and were used as part of the 
sky-subtraction procedure. In addition, a new calibration file for 
the observing mode with 
the Hbb filter and 0\farcs1 scale was taken at a similar operating
temperature to ensure a clean spectral extraction process.

\subsubsection{Reductions}

The data were reduced using the OSIRIS data-reduction pipeline
software \citep{krabbe04}.  In summary, the reduction steps included dark subtraction,
cosmic-ray cleaning and glitch identification, extraction of the
spectra into a data cube, wavelength calibration, sky subtraction
using the offset sky exposures, correction for atmospheric dispersion,
and telluric correction using the extracted spectra of the A0V stars.
The OSIRIS pipeline includes an implementation of the scaled 
sky-subtraction algorithm described by \citet{davies07}, which provides a
more accurate subtraction of the strong $H$-band airglow lines than a
direct subtraction of the offset-sky exposure. However, the current
implementation of the scaled sky routine left noticeable residuals
around the wavelengths of the strongest blended OH features;
we further discuss these sky residuals below.  Following the sky
subtraction and telluric correction, individual exposures were aligned
and co-added to create a final data cube for each galaxy.

We extracted 1-dimensional spectra from each data cube using
rectangular extraction regions.  While we had originally planned to
extract an annular region around the nucleus in order to avoid the
central, AGN-dominated region, we found that in the reduced data the
$H$-band stellar absorption features were visible even in the central
region of the galaxy, and it did not prove necessary to exclude the
central region from the spectral extractions.  We experimented with a
variety of different extraction-region sizes for each galaxy.  Since
the galaxy-light profiles were very centrally concentrated, the
extracted spectra did not change significantly beyond some extraction
aperture size, and in the end we used extraction apertures that were
large enough to contain the majority of the detected galaxy light
without including unnecessarily large amounts of noise from the
surrounding regions.  The extraction aperture sizes were
$1\farcs4\times2\farcs4$, $1\farcs3\times3\farcs4$, and
$1\farcs3\times3\farcs4$, respectively, for Mrk 142, Mrk 766, and NGC
4748.  One-dimensional spectra were similarly extracted from the
reduced data cubes for the velocity template stars.

As a result of the imperfect sky subtraction mentioned previously,
spectral regions around the wavelengths of the brightest OH emission
blends were contaminated by noticeable residuals that appeared to
affect all pixels on the IFU in a similar fashion.  In order to
correct for these additive residuals, for each galaxy we extracted
spectra from rectangular regions at the outer edges of the IFU where
there was very little galaxy continuum detected, with the same number
of pixels as the nuclear extraction, and then subtracted these nearly
blank-sky spectra from the nuclear spectra.  This correction procedure
effectively removed the night-sky emission residuals from the galaxy
spectra.  The procedure was applied for NGC 4748 and Mrk
142, but was not necessary for Mrk 766 since the OH emission-line
strengths appeared to be much more stable during the Mrk 766
observations, leading to a cleaner sky subtraction.

\section{Stellar Velocity Dispersions}

\begin{figure}
\epsscale{1.0}
\plotone{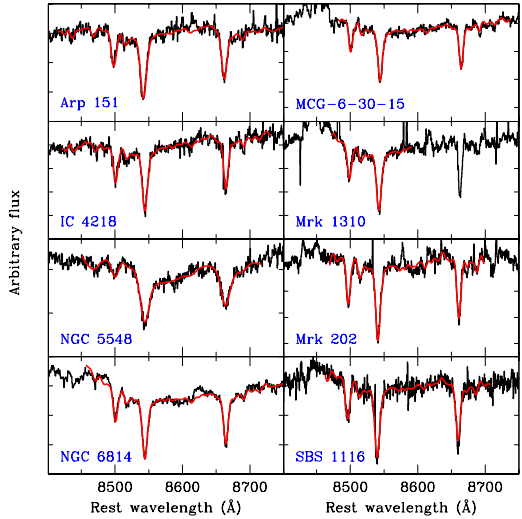}
\caption{Velocity dispersion measurements using the Palomar DBSP and
  the Keck ESI spectra.  The region including the main stellar
  features around the \ion{Ca}{2} triplet is shown together with the
  best-fit template (thick red line).}
\label{fig_dis}       
\end{figure}

To measure stellar velocity dispersions, we used the \ion{Ca}{2} triplet
region (rest frame $\sim$8500\AA) in the optical spectra, and
\ion{Mg}{1} and the CO line region (rest frame $\sim$1.5--1.6 $\micron$) in
the near-IR spectra. Velocity dispersions were determined by comparing
in pixel space the observed galaxy spectra with stellar templates
convolved with a kernel representing Gaussian broadening. Velocity
template stars of various spectral types were observed with the same
instrumental setup at each telescope, minimizing any systematic errors
due to the different spectral resolution of various observing
modes. The minimum $\chi^{2}$ fit was performed using the
Gauss-Hermite Pixel Fitting software (van der Marel 1994).  The merit
of fitting in pixel space as opposed to Fourier space is that typical
AGN narrow emission lines and any residuals from night-sky subtraction
can be easily masked out (e.g., Treu et al. 2004; Woo et al. 2005;
2006), and the quality of the fit can be directly evaluated.
We used low-order polynomials (order 2--4) to model the overall shape of 
the continuum. Extensive and careful comparison was performed using 
continua with various polynomials, and the measured velocity dispersions 
based on each continuum fit with various order polynomials
are consistent within the errors (see e.g., Woo et al. 2004; 2005; 2006). Thus,
continuum fitting with low-order polynomials does not significantly affect the 
velocity dispersion measurements.

\subsection{Optical Data}

For galaxies with optical spectra we mainly used the strong
\ion{Ca}{2} triplet lines (8498, 8542, 8662 \AA) to measure the
stellar velocity dispersions.  We used all velocity template stars
with various spectral types (G8--K5 giants) observed with the same
instrumental setups, in order to account for possible template
mismatches. Fits were performed with each template and the mean of
these measurements was taken as the final velocity dispersion
measurement. The rms scatter around the mean of individual
measurements was added in quadrature to the mean of the individual
measurement errors from the $\chi^{2}$ fit.  Given the high S/N of the
observed spectra, the uncertainties are typically $\sim$5\%.

\begin{figure}
\epsscale{1.0}
\plotone{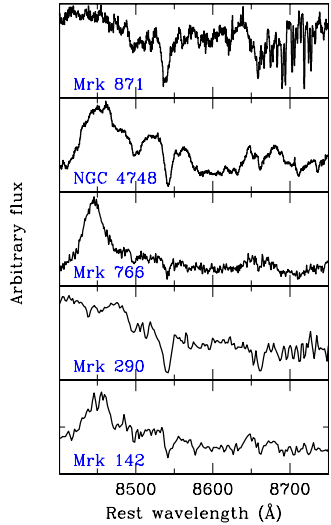}
\caption{Galaxies with contaminated stellar lines. These galaxies
  clearly show the \ion{Ca}{2} triplet lines; however, due to AGN
  contamination, the line profiles are asymmetric. Also, the presence
  of various AGN emission features complicates continuum
  subtraction. The spectra for the first three objects (Mrk 871, NGC
  4748, Mrk 766) are from Palomar DBSP data. The data for Mrk 290 data
  were obtained at the Lick telescope, and Mrk 142 data are from
  SDSS.}
\label{fig_emission}       
\end{figure}
We were able to measure the stellar velocity dispersion of eight objects:
SBS 1116+538A, Arp 151, Mrk 1310, Mrk 202, IC 4218, NGC 5548, NGC
6814, and MCG--6-30-15. Among these, IC 4218 and MCG--6-30-15 were
excluded in the \msigma\ relation analysis since we were not able to
measure the reverberation time scales (Bentz et al. 2009b).  Figure 1
shows examples of the best-fit template along with the observed galaxy
spectra.  Our measurements based on high-quality data are generally
consistent within the errors with previous measurements from the
literature for these galaxies, although many of the previously
existing measurements had very large uncertainties (see
\S\ref{previous} below).

In the case of the other five objects (Mrk 871, NGC 4748, Mrk 766, Mrk 290, 
and Mrk 142), the \ion{Ca}{2} triplet lines were contaminated by AGN emission
lines as shown in Figure 2; hence, we did not obtain reliable stellar
velocity dispersion measurements. Three of these objects, that have 
black hole mass measurements from reverberation mapping, were
observed in the near-IR, and we determined stellar velocity dispersions
for two of them as described in the next section.

\subsection{Infrared Data}

\begin{figure}
\epsscale{1.0}
\plotone{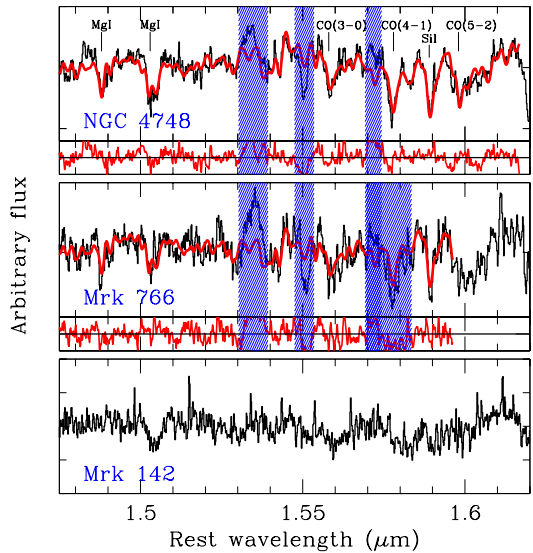}
\caption{Velocity dispersion measurements using the OSIRIS
  spectra. The region including the main stellar features is shown
  together with the best-fit template (thick red line). Individual
  stellar lines are identified in the first panel.  The regions around
  narrow AGN emission lines and regions of severe template mismatch,
  identified by shaded bands, are masked out before fitting. 
  Residuals from the fit are shown at the bottome of NGC 4748 and Mrk 766.}
\label{fig_osiris}       
\end{figure}

Because of the difficulty of measuring the stellar velocity dispersion
in the optical, we obtained near-IR spectra for three galaxies with
reverberation black hole masses (NGC 4748, Mrk 766, Mrk 142) using
OSIRIS.  We also observed several velocity template stars with K and M
spectral types.  As noted by other studies (e.g., Watson et al. 2008),
M giant stars give better fits compared to K giants since M giants are
the representative stellar population in the near-IR bands. After
experimenting with various spectral type velocity templates, we
decided to use M1III and M2III stars for velocity dispersion
measurements and took the mean as the final measurement with an
uncertainty given by the quadrature sum of the fitting error and the
rms scatter of results obtained from different templates.  Figure 3
shows the observed galaxy spectra compared with the best-fit
template. NGC 4748 shows a relatively good fit, while Mrk 766 presents
stronger residuals due to template mismatch. All major $H$-band
stellar lines, such as \ion{Mg}{1} 1.488 \micron, \ion{Mg}{1} 1.503
\micron, CO(3-0) 1.558 \micron, CO(4-1) 1.578 \micron, \ion{Si}{1}
1.589 \micron, and CO(5-2) 1.598 \micron, are present in both objects.
In the case of Mrk 142, which is fainter and at higher redshift than
the other two objects, stellar absorption lines cannot be clearly
identified and we could not obtain a reliable template fit or velocity
dispersion measurement.
 
\subsection{Previous Velocity Dispersion Measurements}
\label{previous}

Several objects in the LAMP sample have preexisting velocity dispersion
measurements in the literature. In most cases the published
results have fairly large uncertainties.  Here we briefly summarize the
previous measurements and describe the available spectroscopic data from 
the Sloan Digital Sky Survey (SDSS).

\smallskip

\textit{SBS 1116+583A:} \citet{gh06} measured $\sigma=50.3\pm18.0$
\kms\ using the SDSS spectrum. However, the velocity dispersion
was not clearly resolved by the SDSS data, which have an instrumental
dispersion of $\sim70$ \kms. 

\smallskip

\textit{Arp 151:} \citet{gh06} measured $\sigma=124\pm12$ \kms\ from the
SDSS spectrum.

\smallskip

\textit{Mrk 1310:} Using the SDSS data, \citet{gh06} reported $\sigma=50\pm16$
\kms, which is below the resolution limit of the SDSS spectrum.

\smallskip

\textit{Mrk 202:} \citet{gh06} measured $\sigma=85.6\pm14$ \kms\ from the SDSS spectrum.

\smallskip

\textit{Mrk 290:} In the SDSS spectrum, the Ca II triplet lines are only weakly
detected, and they appear to be badly contaminated by Ca II emission.
The blue region of the spectrum is too dominated by the AGN for any stellar
features to be clearly visible.

\smallskip

\textit{NGC 4253 (Mrk 766):} \citet{bot05} reported a dispersion of $\sigma=81\pm17$
from the Ca II triplet lines.  However, the Ca II triplet region is moderately
contaminated by emission lines.  Their dispersion measurements were
carried out by cross-correlating the galaxy spectrum with that of a
template star. Without a direct fit of the broadened template star
to the galaxy spectrum, it is not clear how badly the measurement may
be affected by the emission-line contamination.  

An SDSS spectrum of Mrk 766 exists, but it appears to have been taken
with the SDSS fiber positioned about 3\arcsec\ away from the galaxy
nucleus.  Since the SDSS fiber missed the galaxy nucleus, the spectrum
appears similar to that of a Seyfert 2 galaxy, 
dominated by starlight and with very
weak broad components to the Balmer lines.  Our own data from the Lick
monitoring campaign, on the other hand, show a strongly AGN-dominated
spectrum with much more prominent broad lines.  Thus, while the Ca II triplet
and other stellar absorption lines are clearly detected in the SDSS
data, the offset position of the SDSS spectrum means that it should
not be used to infer the bulge velocity dispersion.

\smallskip

\textit{NGC 5548:} The most recent
measurements of the velocity dispersion for this galaxy include
$\sigma=180\pm6$ \kms\ \citep{fer01} and $\sigma=201\pm12$ \kms\
\citep{nelson04}. 

\smallskip

\textit{NGC 6814:}  \citet{nw95} listed a velocity dispersion of
$115\pm18$ \kms.

\smallskip

\textit{Mrk 142}: The SDSS spectrum of Mrk 142 (SDSS
J102531.28+514034.8) shows that the object is strongly AGN dominated,
and the stellar Ca II triplet lines are severely contaminated 
by Ca II emission.
In addition, the spectral region around Mg~$b$ and Fe~5270 is
completely dominated by strong \ion{Fe}{2} emission,
precluding any possibility of measuring the stellar velocity dispersion
in the optical. 

\smallskip

\textit{MCG--6-30-15:} \citet{oliva99} measured $\sigma=159$ \kms\ from
template fits to the CO and Si features in the $H$ band, but they did
not list estimates of the measurement uncertainties.
\citet{mchardy05} used a spectrum around the Ca II triplet to measure $\sigma=93.5\pm8.5$
\kms, using a cross-correlation technique.

\medskip

All of the previous measurements with sufficiently high instrumental resolution
are consistent with our new measurements within the errors, except for
one object, MCG--6-30-15. The previously reported \ion{Ca}{2}-based
velocity dispersion and the $H$-band velocity dispersion of MCG--6-30-15
are dramatically different, and our measured dispersion based on the
\ion{Ca}{2} triplet is significantly lower than either previous
result. Unfortunately, we cannot determine the origin of the
discrepancy other than noting that the \ion{Ca}{2} dispersion from
\cite{mchardy05} was determined using a cross-correlation method, and
therefore it is not easy to directly compare the quality of the result
with our measurement.  As an additional check, we carried out an
independent measurement of \sigmastar\ from the Palomar DBSP data
using the template-fitting code of \citet{bhs02}.  This gave
$\sigmastar=75\pm5$ \kms, consistent with the value of $76\pm3$ \kms\
measured with the \citet{marel94} code.  Since MCG--6-30-15 is not
included in the \msigma\ relation analysis in the next section, this
does not affect our main conclusions.

\begin{deluxetable*}{lcccccc}
\tablecaption{Black hole masses and stellar velocity dispersions} 
\tablehead{ \colhead{Galaxy} &
\colhead{VP} & \colhead{VP ref.} & \colhead{$\sigma_{*}$} & \colhead{error}  & \colhead{$\sigma_*$ ref.}  & \colhead{$\log$ \mbh/\msun} \\ 
\colhead{} & \colhead{($\sigma_{\rm line}^2 R_{\rm BLR}/G$)} & \colhead{} &  \colhead{} & \colhead{} & \colhead{} \\
\colhead{} & \colhead{$10^{6} \msun$} & \colhead{} &  \colhead{\kms} & \colhead{} & \colhead{} 
} 
\startdata
\multicolumn{7}{c}{LAMP Sample} \\
\hline
Arp 151                    & $1.22^{+0.16}_{-0.22}$ & 5  & 118 & 4  & this work & $6.81\pm0.12$\\
IC 4218                      &    \nodata                                  &     & 93 & 4   & this work &  \nodata\\
MCG--6-30-15             &    \nodata                                &     & 76 & 3      & this work &  \nodata\\
Mrk 1310                  & $0.41^{+0.12}_{-0.13}$ & 5  & 84  & 5  & this work & $6.33\pm0.17$\\
Mrk 142                     & $0.40^{+0.12}_{-0.14}$ & 5   & \nodata   &  \nodata     &     & $6.32\pm0.17$ \\
Mrk 202                     & $0.26^{+0.13}_{-0.10}$ & 5  & 78  & 3  & this work & $6.13\pm0.22$\\
NGC 4253 (Mrk 766) & $0.32^{+0.21}_{-0.20}$ & 5 &  93 & 32 & this work & $6.23\pm0.30$\\
NGC 4748                  & $0.47^{+0.16}_{-0.21}$ & 5   &105 &13   & this work & $6.39\pm0.20$\\
NGC 5548                 & $11.9^{+0.46}_{-0.46}$ & 6   & 195 &13  & this work & $7.80\pm0.10$\\
NGC 6814                 & $3.36^{+0.54}_{-0.56}$ & 5   & 95 & 3   & this work & $7.25\pm0.12$\\
SBS 1116+583A      & $1.05^{+0.33}_{-0.29}$ & 5 &  92 & 4  & this work & $6.74\pm0.16$\\
\hline
\multicolumn{7}{c}{Previous Reverberation Sample} \\
\hline
Ark 120             & $27.2^{+3.5}_{-3.5}$ & 1  & 221 & 17 & 1 & $8.15\pm0.11$ \\
3C 120              & $10.1^{+5.7}_ {-4.1}$  & 1  & 162 & 20 & 2 & $7.72\pm0.23$\\
3C 390.3           & $52^{+11.7}_{-11.7}$   & 1  & 273 & 16 & 1 & $8.44\pm0.14$ \\
MRK 79            & $9.52^{+2.61}_{-2.61}$ & 1  & 130 & 12 & 1 & $7.70\pm0.16$ \\
MRK 110          & $4.57^{+1.1}_{-1.1}$   & 1  &  91 &  7 & 3 & $7.38\pm0.14$\\
MRK 279          & $6.35^{+1.67}_{-1.67}$ & 1  & 197 & 12 & 1 & $7.52\pm0.15$ \\
MRK 590          & $8.64^{+1.34}_{-1.34}$ & 1  & 189 &  6 & 1  & $7.66\pm0.12$\\
MRK 871          & $8.98^{+1.4}_{-1.4}$   & 1  & 120 & 15 & 1 & $7.67\pm0.12$ \\
NGC 3227        & $7.67^{+3.9}_{-3.9}$   & 1  & 136 &  4 & 1 & $7.60\pm0.24$\\
NGC 3516        & $7.76^{+2.65}_{-2.65}$ & 1  & 181 &  5 & 1 & $7.61\pm 0.18$\\
NGC 3783        & $5.42^{+0.99}_{-0.99}$ & 1  &  95 & 10 & 4 & $7.45\pm0.13$\\
NGC 4051          & $.287^{+0.09}_{-0.12}$  & 2  &  89 &  3 & 1 & $6.18\pm 0.19$ \\
NGC 4151           & $8.31^{+1.04}_{-0.85}$ & 3  &  97 &  3 & 1 & $7.64\pm 0.11$\\
NGC 4593           & $1.78^{+0.38}_{-0.38}$ & 4  & 135 &  6 & 1 & $6.97\pm0.14$ \\
NGC 7469           & $2.21^{+0.25}_{-0.25}$ & 1  & 131 &  5 & 1 & $7.06\pm0.11$\\
PG 1426+215        & $236^{+70}_{-70}$  & 1  & 217 & 15 & 5 & $9.09\pm 0.16$\\

\label{measurementstable}
\tablecomments{
Col. (1) name.
Col. (2) virial product (\mbh=$f\times$ VP) based on the line dispersion ($\sigma_{\rm line}$) from reverberation mapping.
Col. (3) reference for virial product. 1. Peterson et al.\ 2004;
2. Denney et al.\ 2009; 3. Bentz et al.\ 2006; 4. Denney et al.\ 2006;
5. Bentz et al.\ 2009b; 6. weighted mean of Bentz et al.\ 2007 and Bentz
et al.\ 2009b.
Col. (4) stellar velocity dispersion.
Col. (5) error of stellar velocity dispersion.
Col. (6) reference for stellar velocity dispersion. 1. Nelson et al. 2004;
2. Nelson \& Whittle 1995; 3. Ferrarese et al. 2001; 4. Onken et
al. 2004; 5. Watson et al. 2008.
%; 6. this work.
%
Col. (7) black hole mass calculated assuming the virial coefficient,
$\log f=0.72\pm0.10$. The uncertainty in the black hole mass is
calculated by adding in quadrature the measurement uncertainty of the
virial product in logarithmic space and the uncertainty in the virial coefficient (0.1
dex) as $\rm \epsilon \log \mbh = \sqrt {\rm \epsilon VP^{2} / (VP ~ ln 10)^{2}
 + 0.1^{2}}$.
The average of the positive and negative errors on the virial product 
is taken as the symmetric error ($\epsilon \rm VP$).
}
\end{deluxetable*}

\section{The \msigma\ relation of active galaxies}

The \msigma\ relation is expressed as
\begin{equation}
\log (M_{\rm BH} / M_{\odot}) = \alpha + \beta \log (\sigma_{\ast} / {200\ \mathrm{\kms}}) .
\label{eq:msigma}
\end{equation}
For a sample of quiescent galaxies for which the sphere of influence
of the black hole is spatially resolved in stellar or gas kinematics
studies, Ferrarese \& Ford (2005) measured $\alpha=8.22 \pm 0.06 $ and
$\beta=4.86 \pm 0.43$ with no evidence for intrinsic scatter. Using a
larger sample, including late-type galaxies, G\"ultekin et al.\ (2009)
recently reported $\alpha=8.12\pm0.08$ and $\beta=4.24\pm0.41$ with an
intrinsic scatter $\sigma_{\rm int}=0.44\pm0.06$.

In this section, we investigate the \msigma\ relation of active
galaxies, by combining our 8 new velocity measurements of the LAMP
sample with 16 existing measurements of reverberation 
black hole mass and velocity
dispersion taken from the literature. The relevant properties of the
resulting sample of 24 AGNs are listed in Table~2, along with the
original references for the measurements. 
Note that the selection bias investigated by Lauer et al. (2007) is not
relevant to these reverberation objects since the bias mainly affects black holes in high-mass galaxies where the host
galaxy luminosity function is steeply falling.

As a first step we determine
the slope of the \msigma\ relation (\S 4.1)
assuming that the virial coefficient $f$ is unknown but does not
depend on black hole mass. We also determine the intrinsic scatter in
the relation.  Then in \S~4.2 we determine the mean value and
intrinsic scatter of the virial coefficient $f$ by assuming that the
slope and zero point of the \msigma\ relation are the same for active
and quiescent galaxies. Note that in the following analysis the virial
coefficient $f$ and the zero point $\alpha$ are degenerate; therefore,
additional information is needed to determine what fraction of the
intrinsic scatter is due to $f$ and how much is due to $\alpha$, as we
discuss below.

\subsection{The Slope}

Active galaxies appear to obey an \msigma\ relation with a slope
consistent with that of quiescent galaxies (e.g., Gebhardt et al.\ 2000;
Ferrarese et al.\ 2001). However, the slope has not been well
determined so far, because of limitations in sample size and in the
dynamic range of \mbh\ for active galaxies with reverberation and
stellar velocity dispersion measurements.  The local AGN samples with
single-epoch black hole masses selected from the SDSS
are much larger than the reverberation sample (e.g., Greene \&
Ho 2006; Woo et al. 2008), but these mass estimates have larger
uncertainties ($\sim0.5$ dex) and may suffer from unknown systematic
errors (Collin et al. 2006).

By adding our LAMP galaxies to the previous reverberation sample, we
significantly increase the sample size, particularly at low mass
scales (\mbh\ $< 10^{7} \msun$), and extend the dynamic range to almost
3 decades in \mbh. This enables substantial progress in determining
the slope of the \msigma\ relation, independently for active galaxies.
One important caveat is that the virial coefficient $f$ may vary
across the sample.  Unfortunately, with existing data we cannot
constrain the virial coefficient of each object. Therefore, in the
analysis presented here we will assume that $f$ is
independent of black hole mass, although we allow for intrinsic
scatter at fixed black hole mass. In the next section we will take the
opposite viewpoint and assume that the slope is known from quiescent
samples when we investigate the intrinsic scatter in $f$ and its
dependence on \mbh\ and AGN properties.

In practice, our observable is the so-called virial product VP$\equiv
\sigma_{\rm line}^2 R_\mathrm{BLR}/G$, which is assumed to be related to the
black hole mass by a constant virial coefficient $M_{\rm
BH}$=$f\times$VP. Substituting into Equation~\ref{eq:msigma}, one obtains
\begin{equation}
\log {\rm VP} = \alpha + \beta\ \log\left (\frac{\sigma_{\ast}}{200\ \mathrm{\kms}} \right ) - \log f.
\label{eq:VPsigma}
\end{equation}
In order to determine the slope $\beta$ of the relation, we compute
the posterior distribution function $P$. Assuming uniform priors on
the parameters $\alpha'=\alpha - \log f$, $\beta$, and intrinsic
scatter $\sigma_{\rm int}$, the posterior is equal to the likelihood
given by

\begin{equation}
-2\ln P \equiv \sum_{i=1}^{N} \frac{\left [\log \mathrm{VP}_i - \alpha' - \beta
 \log (\frac{\sigma_{\ast}}{200})_i \right ]^2}{\epsilon_{{\rm
 tot},i}^2} + 2 \ln \sqrt{2\pi \epsilon_{{\rm tot},i}^2},
\label{eq:likeli}
\end{equation}

\noindent
where

\begin{equation}
\epsilon_{{\rm tot},i}^2 \equiv \epsilon_{\log \mathrm{VP}_i}^2 + \beta^{\,2} \epsilon_{\log
 \sigma_*i}^2 + \sigma_{\rm int}^2,
\end{equation}

\noindent
and $\epsilon_{\log \mathrm{VP}}$ is the error on the logarithm of the virial product
($= \rm \epsilon_{\rm VP} / (\rm VP ~ ln 10))$ using the average of the positive and negative errors 
on the virial product as $\rm \epsilon_{\rm VP}$, 
$\rm \epsilon_{\rm \log \sigma_*}$ is the error on the logarithm of the stellar velocity
dispersion ($= \rm \epsilon_{\sigma_*} / (\rm \sigma_* ~ ln 10$)), 
and $\rm \sigma_{\rm int}$ is the intrinsic scatter, which is
treated as a free parameter. 
Note that our adopted likelihood is
equivalent to that adopted as default by G\"ultekin et al.\ (2009) for
their maximum likelihood analysis (see their Appendix A.2), with the
inclusion of an additional Gaussian term to describe the uncertainties
in velocity dispersion. After marginalizing over the other variables,
the one-dimensional posterior distribution functions yield the
following estimates of the free parameters: $\alpha'=7.28\pm0.14$,
$\beta= 3.55 \pm 0.60$, and $\sigma_{\rm int}=0.43\pm 0.08$.

To facilitate comparison with earlier studies (e.g., Tremaine et
al. 2002), we also determine the slope with a ``normalized''
$\chi^{2}$ estimator defined as

\begin{equation}
\chi^{2} \equiv \sum_{i=1}^{N} \frac{\left [\log \mathrm{VP}_i - \alpha' - \beta
 \log (\frac{\sigma_{\ast}}{200})_i \right ]^2}{\epsilon_{{\rm
 tot},i}^2},
\end{equation}
where the intrinsic scatter $\sigma_{\rm int}=0.43$ is set to yield
$\chi^{2}$ per degree of freedom equal to unity. The best estimate of
the slope based on this normalized $\chi^{2}$ estimator is
$\beta=3.72^{+0.40}_{-0.39}$. If we assume no intrinsic scatter, then
the best-fit slope becomes slightly steeper as
$\beta=3.97^{+0.13}_{-0.13}$. However, confirming our previous result
based on a full posterior analysis, the intrinsic scatter cannot be
negligible since the $\chi^{2}$ per degree of freedom would be much larger
than unity ($\sim$14) if the intrinsic scatter is assumed to be zero.

These results indicate that the slopes of the \msigma\ relation for
active and quiescent galaxies are consistent within the
uncertainties. The total intrinsic scatter is $0.43\pm0.08$ dex, which
is a combination of the scatter of the \msigma\ relation and that of
the virial coefficient.  Without additional information, we cannot
separate the contribution of the intrinsic scatter on the virial
coefficient (i.e. the diversity of broad line region kinematic
structure amongst AGN) from the intrinsic scatter in black hole mass
at fixed velocity dispersion (i.e., the diversity of black hole masses
amongst galaxies). We will address this issue in \S 4.2.

\subsection{The Virial Coefficient}

In order to transform the virial product into an actual black hole
mass, we need to know the value of the virial coefficient $f$. Our
current understanding of the geometry and kinematics of the BLR is
insufficient to compute $f$ from first principles. This creates a
systematic uncertainty in the determination of black hole mass from
reverberation mapping.  As an example, if a spherical isotropic
velocity distribution is assumed, then the velocity dispersion of the
gas is a factor of $\sqrt{3}$ larger than the measured line-of-sight
velocity dispersion ($\sigma_{\rm line}$), hence a value of $f=3$ is
often assumed for a spherically symmetric BLR
\citep[e.g.,][]{wandel99, kaspi00}.  If, instead, the BLR is described
by a circular rotating disk, then the value of $f$ can be 
a few times larger than the isotropic case, depending on the
inclination angle and the scale height of the disk (see Collin et al.\
2006 for more details).

As mentioned above, determining the virial coefficient for individual
AGNs is not feasible due to the limited spatial resolution of the
current and foreseeable future instruments since the angular size of
the BLR is microarcseconds.  Therefore, in practice an average virial
coefficient is usually determined by assuming that local Seyfert
galaxies and quasar host galaxies follow the same \msigma\ relation
(Onken et al.\ 2004) and deriving the appropriate value for $f$.
Using this method, Onken et al. (2004) measured $\langle f
\rangle=5.5\pm1.8$ based on a sample of 14 reverberation-mapped
Seyfert galaxies with measured stellar velocity dispersions.

In this section we apply the same methodology to our enlarged sample
to determine the average value of $f$ as well as the intrinsic
scatter, assuming the intercept ($\alpha$) and slope ($\beta$) of the
\msigma\ relation for quiescent galaxies. In practice, we modify the
likelihood given in Equation~\ref{eq:likeli} by fixing the slope and
intercept to the values obtained by two independent groups, Ferrarese
\& Ford (2005) and G\"ultekin et al. (2009).  Figures 4 and 5 show the
best-fit \msigma\ relations for the enlarged reverberation sample,
respectively assuming the quiescent \msigma\ relation from Ferrarese
\& Ford (2005) and that from G\"ultekin et al. (2009).  Adopting the
\msigma\ relation of Ferrarese \& Ford (2005; $\alpha=8.22$ and
$\beta=4.86$), we determine $\log f=0.71 \pm 0.10$ and $\sigma_{\rm int} =
0.46^{+0.07}_{-0.09}$ dex.  Adopting the \msigma\ relation from
G\"ultekin et al (2009; $\alpha=8.12$ and $\beta=4.24$), we obtain
$\log f=0.72^{+0.09}_{-0.10}$ and $\sigma_{\rm int}= 0.44\pm0.07$.  The
average values are in good agreement with $f=5.5\pm1.8$ (i.e., $\log
f=0.74^{+0.10}_{-0.13}$) as found by Onken et al.\ (2004). The average
value is inconsistent with that expected for a spherical BLR ($f=3$),
and closer to that expected for a disk-like geometry, although there
may be a diversity of geometries and large-scale kinematics (e.g.,
Bentz et al.\ 2009b; Denney et al.\ 2009).

\begin{figure}
\epsscale{1.8}
\plottwo{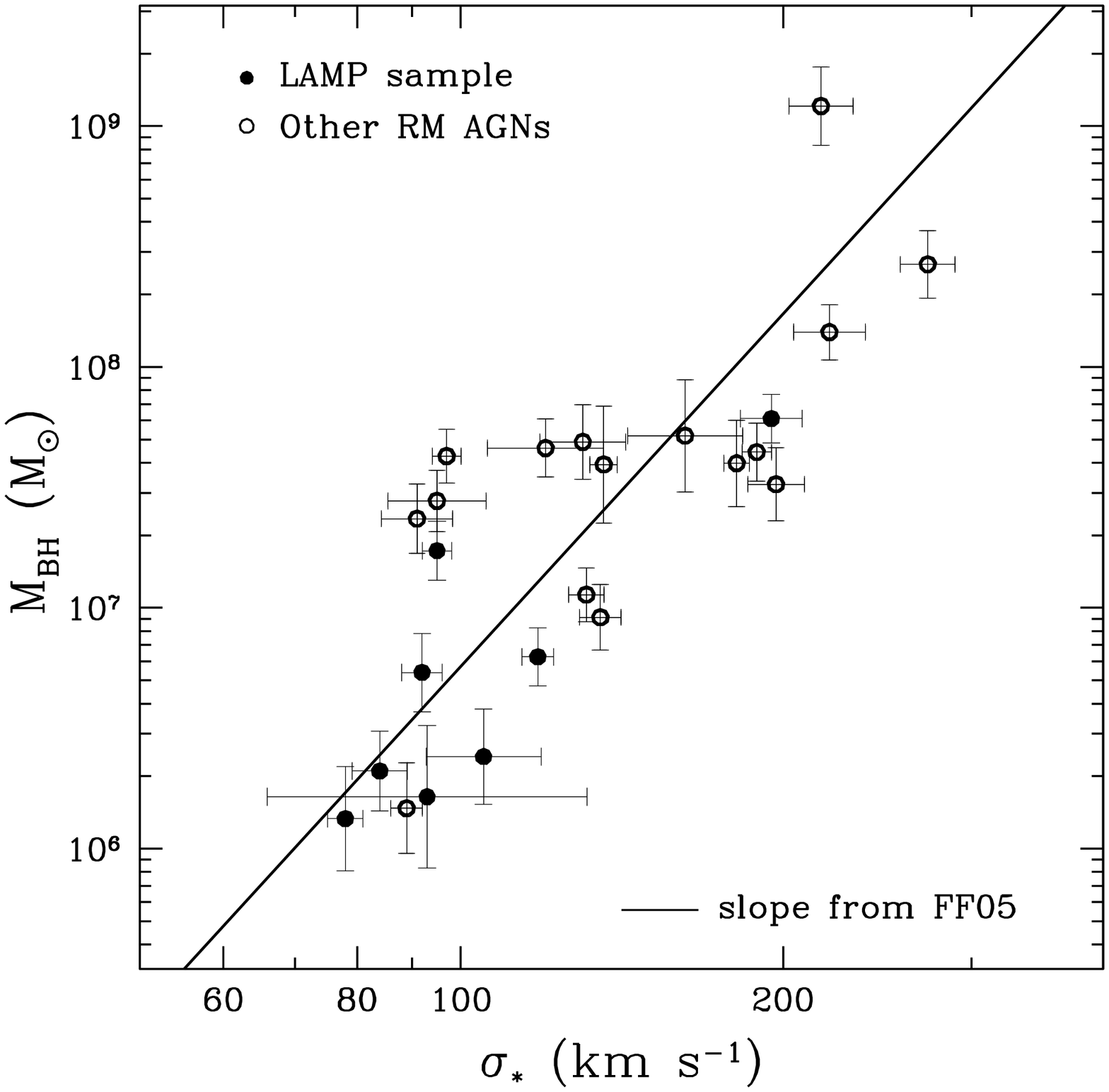}{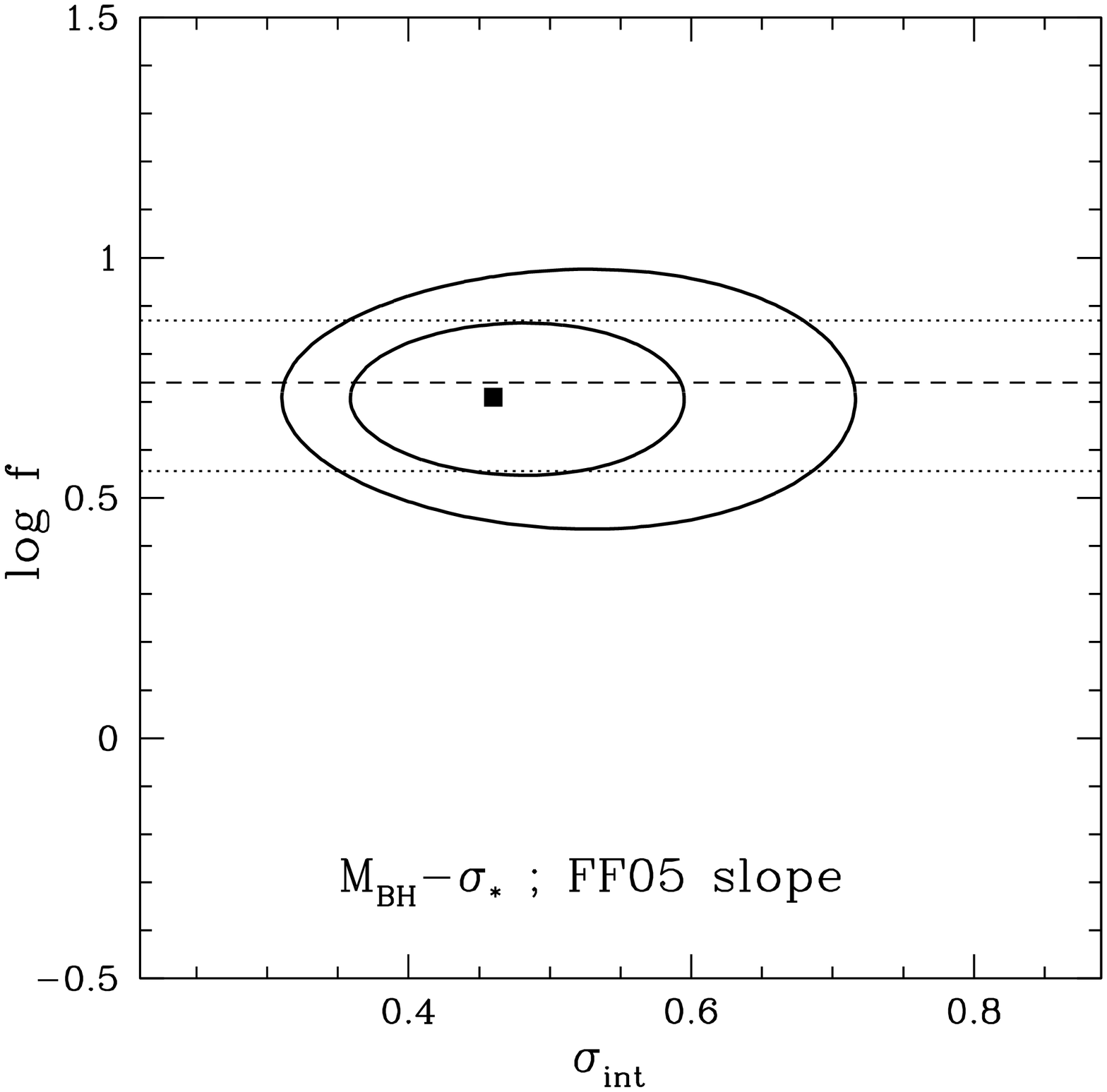}
\caption{{\it Top:} \msigma\ relation of the combined reverberation sample,
  adopting the slope from Ferrarese \& Ford (2005) and the best-fit
  virial coefficient from this work. {\it Bottom:} Posterior probability
  contour levels (68\% and 95\%) for the virial coefficient $f$ and
  its intrinsic scatter, as determined assuming the slope from
  Ferrarese \& Ford (2005). The horizontal lines represent the
  previous estimate from Onken et al. (2004; dashed line) with its
  68\% uncertainties (dotted lines).}
\label{fig:FF05}
\end{figure}

As discussed at the end of \S~4.1, the intrinsic scatter determined in
this section is a combination of intrinsic scatter in $f$ and in zero
point of the \msigma\ relation (which are degenerate in Equation
4). We need additional information to break this degeneracy. If the
intrinsic scatter of the active galaxy \msigma\ relation were close to
0.31 dex (as determined for elliptical galaxies by G\"ultekin et al.\
2009), then the residual intrinsic scatter of the virial coefficient would be
0.31 dex. In contrast, if the scatter of the \msigma\ relation of
active galaxies were closer to 0.44 dex (as found for all galaxies by
G\"ultekin et al.\ 2009), then the intrinsic scatter in $f$ would be
close to zero. Although we cannot break this degeneracy with current
data, the bottom line is the same as far as determining black hole
masses from reverberation mapping data is concerned (and from 
single-epoch data, as a consequence). Since for the purpose of the
calibration of $f$ the two sources of scatter are indistinguishable,
the uncertainty in our calibration of $f$ (whether it is due to
diversity in BLR physics or diversity in the galaxy-AGN
connection) is the largest remaining uncertainty in black hole mass
determinations that rely on this technique.

\begin{figure}
\epsscale{1.8}
\plottwo{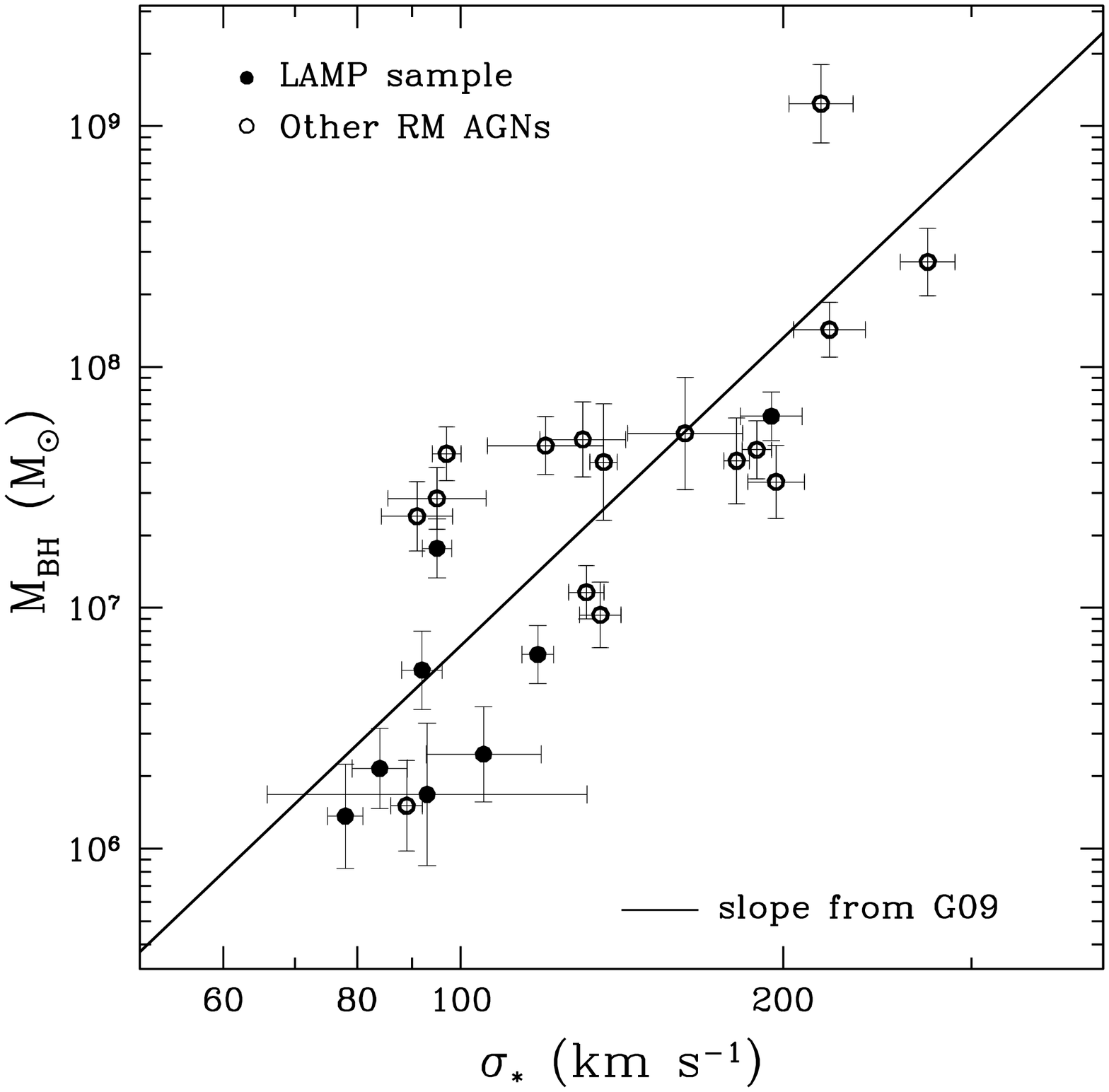}{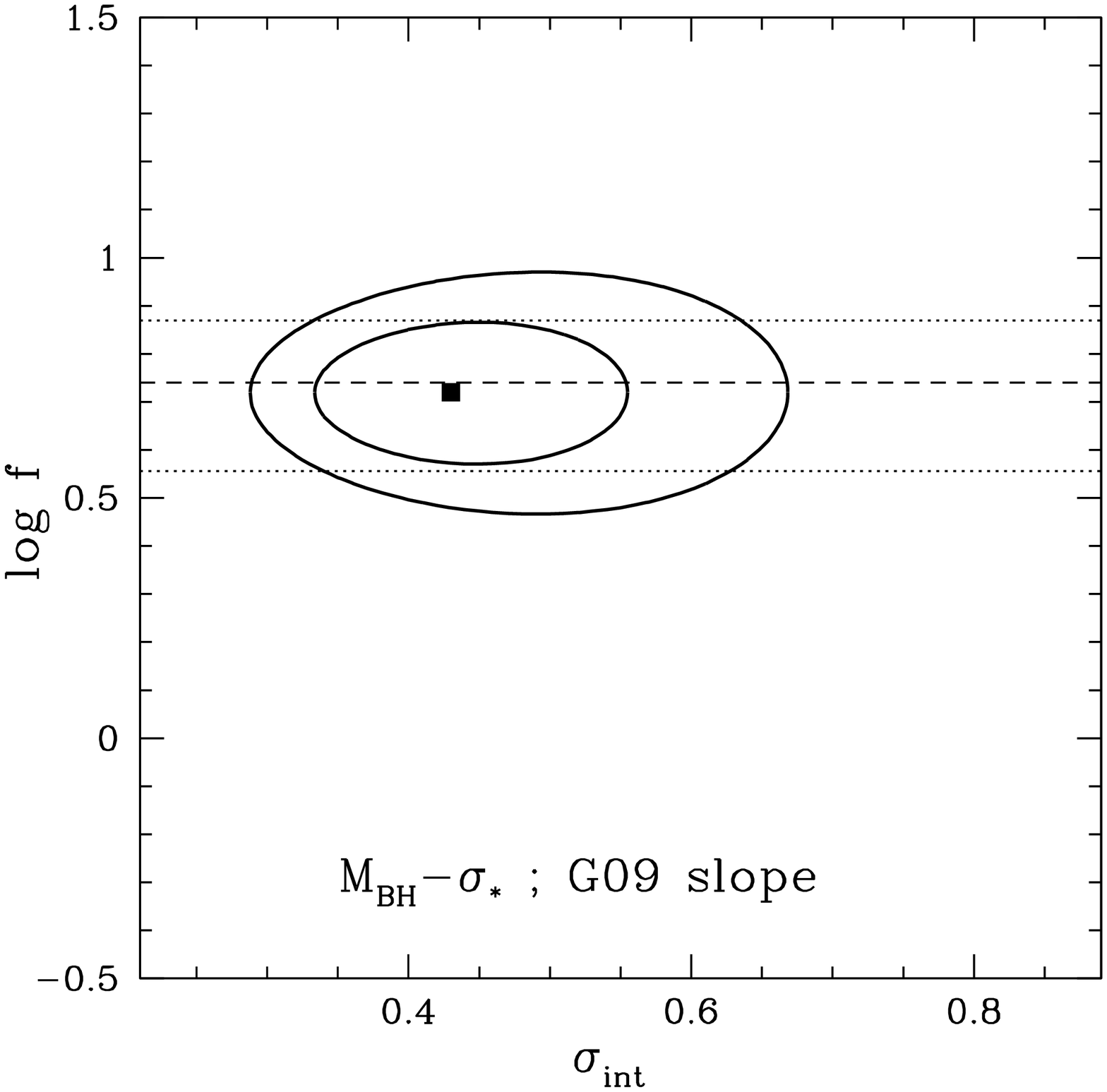}
\caption{As in figure~\ref{fig:FF05}, but adopting the slope from
  G\"ultekin et al.\ (2009).}
\label{fig:G09}       
\end{figure}

To investigate any dependence of the virial coefficient on AGN
properties, we collected from the literature the H$\beta$ line widths
($V_{\rm FWHM}$ and $\sigma_{\rm line}$) and the optical AGN
luminosities for reverberation-mapped AGNs, corrected for the 
host-galaxy contribution (Peterson et al. 2004; Bentz et al. 2009a).  When
multiple luminosity measurements are available for given objects, we
used the weighted mean of the luminosity (Table 9 of Bentz et
al. 2009a).  For the seven LAMP objects, with \mbh\ and velocity
dispersion measurements, {\it Hubble Space Telescope} 
imaging is not yet available for
measuring the starlight correction to the spectroscopic luminosity.
Thus we inferred the AGN luminosity from the measured time lag using
the size-luminosity relation (Bentz et al. 2009a).  The bolometric
luminosity was calculated by multiplying the optical AGN luminosity by
a factor of ten (Woo \& Urry 2002), and we assumed 0.3 dex error in 
the Eddington ratio.

In Figure 6, we plot the residuals from the \msigma\ relation
with $\alpha=8.12$, $\beta=4.24$ taken from G\"ultekin et al.\ (2009), 
and $\log f =0.72$ as determined in \S 6.2.
The residuals were determined as $\Delta \log \mbh = \log VP + \log f
- (\alpha + \beta \log \sigma_*)$.  To test for potential correlations
between residuals and line properties or Eddington ratios, we fit each
data set with a straight line by minimizing the normalized $\chi^2$
estimator.  We find no dependence of the residuals on the line width
(either $V_{\rm FWHM}$ or $\sigma_{\rm line}$).  In the case of the
Eddington ratio, we find a weak dependence with a non-zero slope
$(-2.3^{+0.6}_{-1.1}$), significant at the 95\% level; however this
weak trend is mainly due to one object with the lowest Eddington ratio
in the sample.  When we remove that object, the best fit slope becomes
consistent with no correlation. Further investigation using a larger
sample, evenly distributed over the range of the Eddington ratio, is
required to better constrain the residual dependence on the Eddington
ratio.  In the case of the line width, it is not straightforward to
test the dependence since the line width of most of our objects is
relatively small, $\sigma_{\rm line} < 2000$ \kms.  By dividing our
sample at $\sigma_{\rm line}= 1500$ \kms\ into two groups of similar
sample size, we separately measured the virial coefficient for
narrower-line and broader-line AGNs. The difference in the virial
coefficient is $\Delta \log f = 0.1-0.2$, which is not significant
given the uncertainty of 0.15 dex on the virial coefficient.

\begin{figure}
\epsscale{1.2}
\plotone{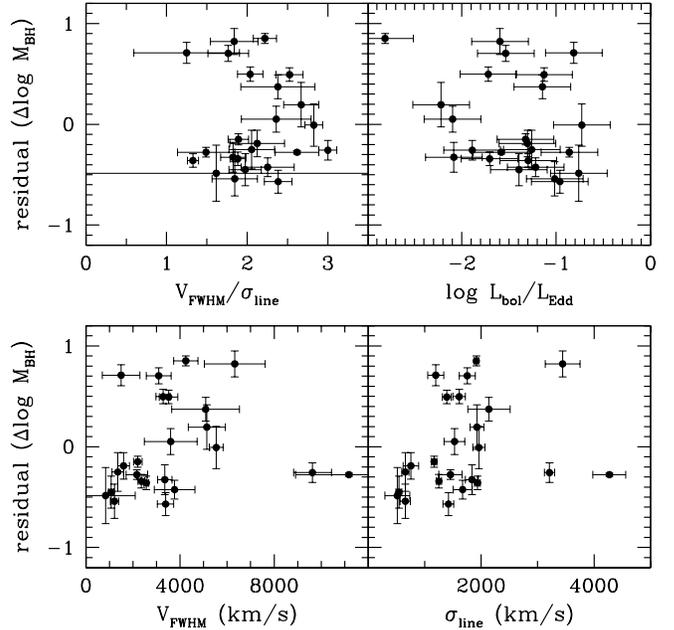}
\caption{Dependence of residuals from the \msigma\ relation ($\Delta
  \log \mbh = (\log \mathrm{VP} + \log f) - (\alpha +\beta \log
  \sigma/(200~ \mathrm{\kms})$) on parameters related to the accretion
  state: $V_{\rm FWHM}/\sigma_{\rm line}$ (top left); Eddington ratio
  (top right); $V_{\rm FWHM}$ (bottom left); line dispersion
  $\sigma_{\rm line}$ (bottom right) of the H$\beta$ line. 
  In this plot we adopt the local relation with $\alpha=8.12$, 
  $\beta=4.24$ taken from G\"ultekin et al.\ (2009), 
  and $\log f =0.72$ as determined in \S 6.2.}
\label{fig:residual_F_g09}       
\end{figure}
\section{Discussion}

\begin{figure*}
\epsscale{1.0}
\plotone{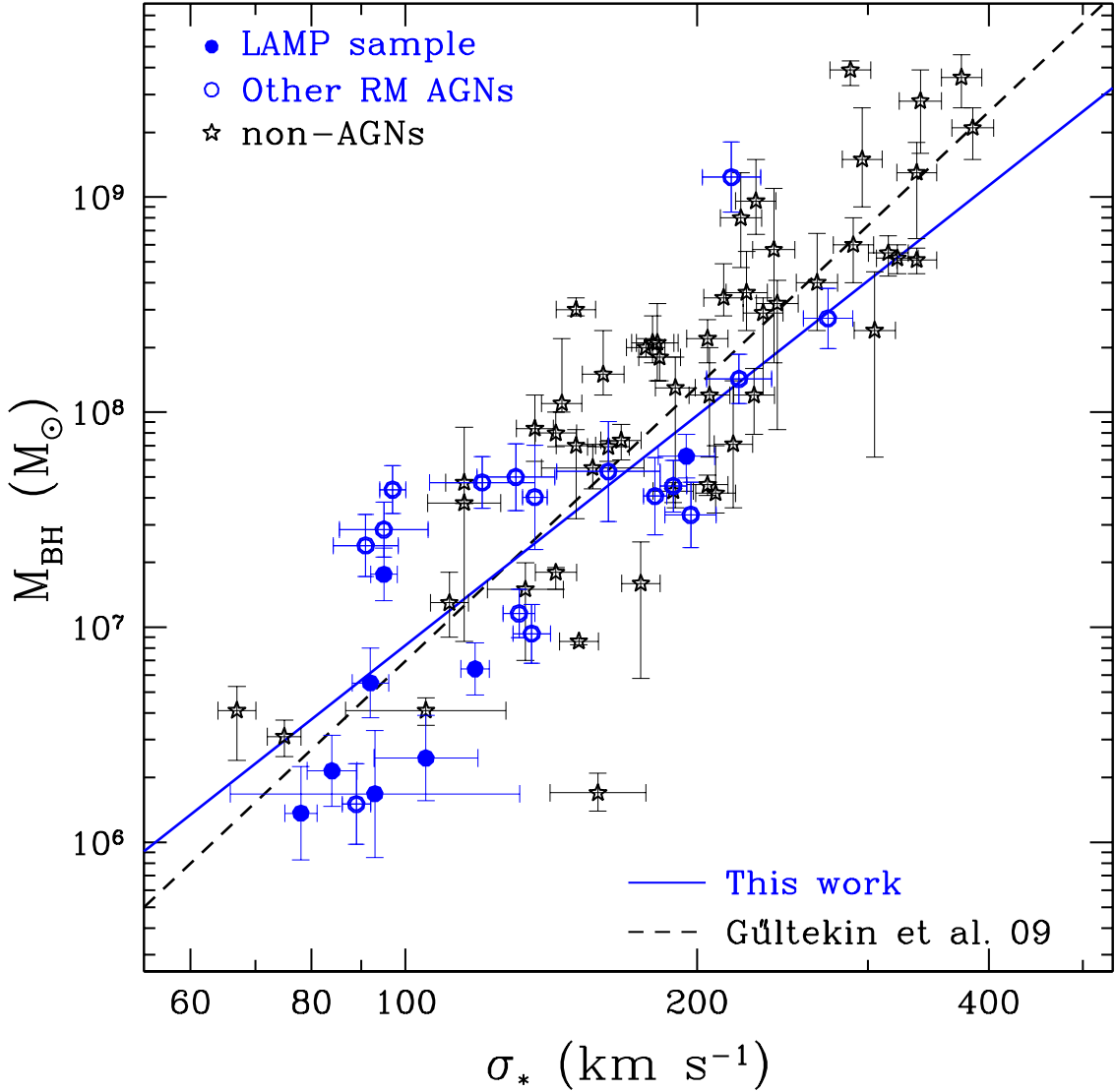}
\caption{The \msigma\ relation of active galaxies with reverberation
  black hole masses (blue), compared with non-active galaxies with
  dynamical black hole masses. The reverberation masses were
  determined assuming the virial coefficient, $\log f=0.72\pm0.10$.
  The solid line is the best-fit slope of the active galaxies while
  the dashed line is the best fit to the inactive
  galaxy samples from G\"ultekin et al. (2009).}
\label{fig_msigma}       
\end{figure*}

We present the \msigma\ relation of the reverberation sample in Figure
7, using the slope, $\beta=3.55\pm0.60$, determined in \S 4.1 and the
average virial coefficient, $\langle\log f\rangle=0.72\pm0.10$,
determined in \S 4.2.
Compared to the local quiescent galaxies, active galaxies follow a
consistent \msigma\ relation with a similar slope and scatter.  Note
that the mean black hole mass of the reverberation sample
($\langle\log \mbh/\msun \rangle=7.3\pm0.74$) is an order of magnitude
smaller than that of quiescent galaxies ($\langle\log
\mbh/\msun \rangle=8.2\pm0.79$).  The slightly shallower slope of the
reverberation sample is consistent with the trend in the quiescent
galaxies that the slope is shallower for galaxies with lower velocity
dispersion ($\sigma_* < 200$ \kms); see G\"ultekin et al. (2009).  In
contrast, the slope of late-type quiescent galaxies ($4.58\pm1.58$)
seems higher than the slope of our active galaxies, which are mainly
late-type galaxies.  However, given the uncertainty in the slope
($\beta=3.55\pm0.60$) of the active galaxy \msigma\ relation, the
difference between quiescent and active galaxies is only marginal. We
did not attempt to divide our sample into various morphology groups or
a few mass bins to test the dependence of the slope, since the sample
size is still small and biased toward lower mass objects.

The smaller mean \mbh\ of the reverberation sample is due to the 
relative difficulty of measuring the stellar velocity dispersions 
for the more luminous and higher redshift quasars that have the 
largest black hole masses.
Although there are 17 quasars with measured reverberation
black hole masses, the stellar velocity dispersions have not been measured
for most of those high-luminosity objects
(see Watson et al. 2008; Dasyra et al. 2007).  In our current sample,
there are only three objects above $3\times 10^{8}$ \msun.  In order
to better constrain the \msigma\ relation of active galaxies, it will
be necessary to pursue further measurements of velocity dispersions
for reverberation-mapped AGNs at high masses.  Additional stellar
velocity dispersion measurements will be available based on Keck
OSIRIS spectra in the near future (Woo \& Malkan, in preparation).

The key assumption in determining the slope of the \msigma\ relation
is that the virial coefficient does not systematically vary as a
function of black hole mass. Currently, we do not have sufficient
information to test whether this assumption is valid. In an empirical
study of the systematic effects of emission-line profiles, Collin et al. (2006) showed that it
was necessary to use a smaller value of the virial coefficient for
AGNs with larger $V_{\rm FWHM}$.  In contrast, the change of the
virial coefficient was not required if the line dispersion
($\sigma_{\rm line}$) was used instead.  These results reflect the
fact that there is a large range of $V_{\rm FWHM}/\sigma_{\rm line}$
ratio, and that systematic uncertainties are higher if $V_{\rm FWHM}$
is used.

With the enlarged sample of 24 AGNs, we investigated the dependence of
the virial coefficient on the line properties and the Eddington
ratio. However, we did not clearly detect any increasing trend of the
intrinsic scatter with the line profile, line width, or Eddington
ratio, implying that the \mbh\ based on the mean virial coefficient is
not systematically over- or under-estimated as a function of line width or
the Eddington ratio.  We note that since our sample is still small and
biased to narrower-line and lower Eddington ratio objects, further
investigation with a larger sample, particularly at higher mass
scales, is required to better constrain the systematic errors.

For single-epoch black hole mass determination, we suggest using $\log f=
0.72\pm0.10$ when the gas velocity is determined from the line dispersion
($\sigma_{\rm line}$). 
If the FWHM of an emission line ($V_{\rm FWHM}$) is
measured instead of the line dispersion, then additional
information on the $V_{\rm FWHM}/\sigma_{\rm line}$ ratio is needed in
the black hole mass calculation.
The value of $V_{\rm FWHM}/\sigma_{\rm line}$ is typically
assumed to be 2 (Netzer 1990), although there is a range of
values. For example, the mean of $V_{\rm FWHM}/\sigma_{\rm line}$ of our reverberation
sample is $2.09$ with a standard deviation of $0.45$ (see also Peterson et
al. 2004 and McGill et al.\ 2008). 
We did not attempt to determine an average virial coefficient using
the virial product based on $V_{\rm FWHM}$.  
A correction factor depending on $V_{\rm FWHM}/\sigma_{\rm line}$ should be applied in addition to
the virial coefficient, $\log f= 0.72$, when gas velocity is determined from $V_{\rm FWHM}$.
Our empirical measurement of the mean virial coefficient 
is not very different from that of Onken et al. (2004) although the sample size
and the dynamical range of the reverberation sample is significantly improved.

We note that our results on the mean virial coefficient are based on broad-line 
dispersions measured from the rms spectra,
which are not generally available for single-epoch black hole mass determinations.
Detailed comparison of broad-line profiles and widths between rms and single-epoch spectra
is necessary to  understand the additional uncertainty of single-epoch black hole mass estimates
(see Denney et al. 2009), and we will pursue further investigations of
this issue using LAMP data in the future.

\section{Summary}

We summarize the main results as follows.

\smallskip
(1) Using the high S/N optical and near-IR spectra obtained at
the Keck, Palomar, and Lick Observatories, we measured the stellar
velocity dispersion of 10 local Seyfert galaxies, including 7 objects
(SBS 1116+ 583A, Arp 151, Mrk 1310, Mrk 202, NGC 4253, NGC 4748, NGC
6814) with newly determined reverberation black hole masses from the
LAMP project (and NGC 5548, IC 4218 and MCG--6-30-15).

\smallskip
(2) For a total sample of 24 local AGNs, combining our new
stellar velocity dispersion measurements of the LAMP sample and the
previous reverberation sample with the stellar velocity dispersion
from the literature, we determined the slope and the intrinsic scatter
of the \msigma\ relation in the range of black hole mass
$10^{6}<\mbh/\msun<10^{9}$.  The best-fit slope is
$\beta=3.55\pm0.60$, consistent within the uncertainty with the slope
of quiescent galaxies. The intrinsic scatter, $0.43\pm 0.08$ dex, of
active galaxies is also similar to that of quiescent galaxies. Thus,
we find no evidence for dependence of the present-day \msigma\ relation
slope on the level of activity of the central black hole.

\smallskip
(3) We determined the virial coefficient using the slope and the
intercept of the \msigma\ relation of quiescent galaxies, taken from
two groups (Ferrarese \& Ford 2005; G\"ultekin et al.\ 2009). The
best-fit virial coefficient ($\log f$) is $0.71\pm0.10$
($0.72^{+0.09}_{-0.10}$) with an intrinsic scatter,
$0.46^{+0.07}_{-0.09}$ ($0.44\pm0.07$) with the slope, 4.86 (4.24)
taken from Ferrarese \& Ford (G\"ultekin et al. 2009).  We take
$f=5.2$ (i.e., $\log f=0.72$) as the best value of the mean virial
coefficient, which is a factor of $\sim$1.7 larger than the standard
factor obtained for isotropic spatial and velocity distribution.  The
substantial intrinsic scatter indicates that the virial coefficient is
the main source of uncertainties in determining black hole masses,
either using reverberation mapping or single-epoch spectra.

\acknowledgments

J.H.W. acknowledges support from Seoul National University by the
Research Settlement Fund for new faculty, and support from NASA
through Hubble Fellowship grant HF-51249 awarded by the Space
Telescope Science Institute.  T.T. acknowledges support from the NSF
through CAREER award NSF-0642621, by the Sloan Foundation through a
Sloan Research Fellowship, and by the Packard Foundation through a
Packard Fellowship. Research by A.J.B. and A.V.F. is supported by NSF
grants AST-0548198 and AST-0908886, respectively.  The work of
D.S. was carried out at Jet Propulsion Laboratory, California
Institute of Technology, under a contract with NASA.
Research by G. C. is supported by NSF grant AST-0507450.
We thank the referee for useful suggestions.

Data presented herein were obtained at the W. M. Keck Observatory,
which is operated as a scientific partnership among Caltech, the
University of California, and NASA; it was made possible by the
generous financial support of the W. M. Keck Foundation.  The authors
wish to recognize and acknowledge the very significant cultural role
and reverence that the summit of Mauna Kea has always had within the
indigenous Hawaiian community; we are most fortunate to have the
opportunity to conduct observations from this mountain. We are
grateful for the assistance of the staffs at the Keck, Palomar, and
Lick Observatories. We thank Kartik Sheth for assisting with some of
the Palomar observations.  We acknowledge the usage of the HyperLeda
database (http://leda.univ-lyon1.fr).  This research has made use of
the NASA/IPAC Extragalactic Database (NED) which is operated by the
Jet Propulsion Laboratory, California Institute of Technology, under
contract with the National Aeronautics and Space Administration.

\newpage

\end{document}